# Phase-Field Modeling of Liquid Phases with Ordering

Yanzhou Ji* and Chengyin Wu

*Department of Materials Science and Engineering, The Ohio State University, Columbus OH, 43210, USA*

**Abstract**

A strong negative enthalpy of mixing can lead to ordering in liquid solutions. To describe ordering in liquids, different thermodynamic descriptions of liquid phases have been developed in the CALPHAD approach, including the associate model and the modified quasichemical model. In this study, we show that the key to directly incorporate these liquid phase models and their corresponding CALPHAD-type free energy formulations in phase-field simulations, is to identify the internal processes and to properly define the internal thermodynamic variables. The free energy of the liquid phase can then be converted to a function of component compositions and internal variables, and the evolution of the internal variables will be determined via Allen-Cahn-type relaxation equations. We demonstrate this modeling strategy using the formation of stoichiometric phases from liquid phases in Na-Sb and Na-Sn systems. This study will pave the way for predicting microstructure evolution in liquids with ordering using phase-field simulations.

*Corresponding author. Email: ji.730@osu.edu

## 1. Introduction

The enthalpy of mixing ($\Delta H_{mix}$) critically determines the thermodynamic stability of a liquid solution. For liquid solutions with a negative $\Delta H_{mix}$, there is a tendency for ordering since the unlike interactions (e.g., A-B bonds in a binary system) are energetically more favorable than the like-like interactions (e.g., A-A and B-B bonds). In some of these liquid solutions, localized structures with short-range ordering (SRO) can be observed, which may play a key role in forming long-range-ordered phases during solidification. To elucidate the stability of the SRO structures and the liquid phase, the kinetic pathways of solidification and microstructure evolution in such liquid phases, it is critical to correctly and accurately account for the SRO in the thermodynamic description of the liquid phase.

To date, there are mainly two different ways to account for SRO in liquid phases in the CALPHAD community. One is the associate model, which presumes certain SRO species with fixed stoichiometry, i.e., "associates", and the molar Gibbs free energy of the liquid phase is formulated as a function of the site fractions of the solute, solvent, and associates. The other is the modified quasichemical model in the pair approximation (MQMPA)[1, 2], in which the Gibbs free energy of the liquid phase is formulated as a function of the number of moles of the solute and solvent species, as well as the number of moles of the quasichemical pairs. The associate model is relatively simple, but the assumption of the associates may not be physical and it is not easy to be extended to multicomponent systems. On the other hand, the MQMPA can more accurately describe the SRO behavior and the configuration entropy change due to the quasichemical pairs, and can account for the composition-dependent coordination numbers, but is more complicated to implement. In addition, the ionic two sublattice model has also been applied for ionic systems such as oxides and molten salts, which separates the cation and anion sublattices, and the valence of the charged vacancy, as well as the stoichiometries of the sublattices are allowed to vary. This model can be used in combination with the associate model to describe ordering behaviors in liquid phases of ionic systems.

With these available thermodynamic descriptions of liquid phases with ordering, it is possible to analyze the phase stabilities and phase transformation kinetics using the optimized thermodynamic databases. For example, the phase-field model has been widely applied to predict the microstructure evolution during solidification and precipitation. However, most of the existing phase-field investigations did not account for the possible ordering behaviors of the liquid phases. None of the phase-field investigations directly incorporate the Gibbs free energy formulations of the associate model or the MQMPA without any approximations. A major reason for such a lack of efforts is that these liquid phase models involve extra degrees of freedom than only solute composition itself, making it not straightforward to be directly incorporated in phase-field simulations.

In this work, we will directly incorporate the associate model and the MQMPA of liquid phases in phase-field simulations. To do so, we will first identify the extra degrees of freedom in these models, which correspond to different internal processes within the liquid phase. We then define the internal process order parameters (IPOPs) characterizing the extent of these internal processes, and evolve these IPOPs using Allen-Cahn equations. The microstructure evolution involving the liquid phase can then be simulated using the coupled Allen-Cahn equations for the phase order parameters, Allen-Cahn equations for the IPOPs, and diffusion equations for compositions. The model will be demonstrated using Na-Sb and Na-Sn systems.

## 2. Model description

The key for phase-field modeling of liquid phases with ordering is to identify the internal degrees of freedom related to ordering, define the IPOPs, and eventually evolve the IPOPs using Allen-Cahn equations, as illustrated in Fig. 1. In this section, we will show in detail how this framework can be applied for liquid phases described by the associate model and the MQMPA model.

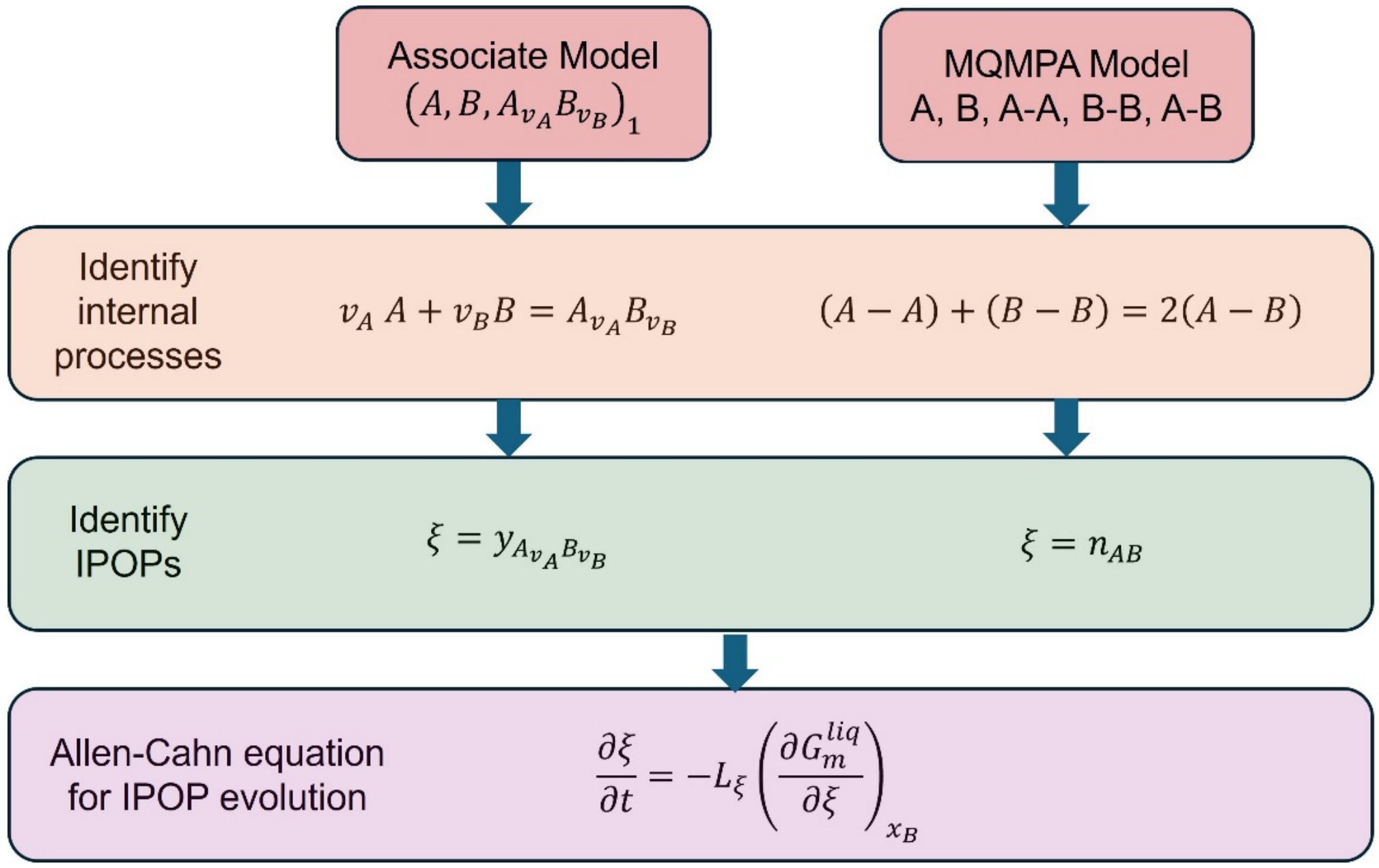

Fig. 1 Illustration of the phase-field model for liquid phases described by the associate model and the MQMPA model.

### 2.1 Phase-field model of individual liquid phases

#### *2.1.1 Phase-field model of liquid phases described by the associate model*

Let us consider a binary liquid with components A and B. Due to the strongly negative $\Delta H_{mix}$, SRO may take place, forming an associate with the chemical formula $A_{\upsilon_A}B_{\upsilon_B}$ where $\upsilon_A$ and $\upsilon_B$ are positive constants. The liquid phase is then described by the sublattice model (A, B, $A_{\upsilon_A}B_{\upsilon_B}$)$_1$. Based on the analysis of [3, 4], this sublattice model has one internal degrees of freedom, corresponding to the internal reaction

| $$\upsilon_A A + \upsilon_B B = A_{\upsilon_A}B_{\upsilon_B}$$ | (R1) |
|---|---|

to form the stoichiometric associate. To describe the extent of this internal reaction, it is natural to choose the site fraction of the associate, $y_{A_{\upsilon_A}B_{\upsilon_B}}$, as the IPOP. The sublattice site fractions $y_A$, $y_B$ and $y_{A_{\upsilon_A}B_{\upsilon_B}}$ can then be converted to the solute composition in the liquid phase, $x_B$, and the IPOP, $\xi$, via (see Supplementary Material S1 of [5]):

| $$y_A + y_B + y_{A_{\upsilon_A}B_{\upsilon_B}} = 1$$ | (1a) |
|---|---|
| $$x_B = \frac{y_B + \upsilon_B y_{A_{\upsilon_A}B_{\upsilon_B}}}{y_A + y_B + (\upsilon_A + \upsilon_B) y_{A_{\upsilon_A}B_{\upsilon_B}}}$$ | (1b) |
| $$\xi = y_{A_{\upsilon_A}B_{\upsilon_B}}$$ | (1c) |

which gives

| $$y_A = 1 - x_B + \xi[-1 + \upsilon_B - (\upsilon_A + \upsilon_B - 1)x_B]$$ | (2a) |
|---|---|
| $$y_B = x_B - \xi\upsilon_B + \xi x_B(\upsilon_A + \upsilon_B - 1)$$ | (2b) |
| $$y_{A_{\upsilon_A}B_{\upsilon_B}} = \xi$$ | (2c) |

The Gibbs free energy of a mole of formula unit of the liquid phase, $G_M^{liq}\left(y_A, y_B, y_{A_{\upsilon_A}B_{\upsilon_B}}\right)$ can then be converted to a mole of atom of the liquid phase, $G_m^{liq}(x_B, \xi)$ via $G_m^{liq} = \frac{G_M^{liq}}{y_A + y_B + (\upsilon_A + \upsilon_B) y_{A_{\upsilon_A}B_{\upsilon_B}}}$. The evolution of $\xi$ is governed by the Allen-Cahn equation:

$$\frac{\partial \xi}{\partial t} = -L_\xi \left(\frac{\partial G_m^{liq}}{\partial \xi}\right)_{x_B} \quad (3)$$

where $L_\xi$ is a kinetic coefficient related to the rate of the internal reaction. The driving force $\left(\frac{\partial G_m^{liq}}{\partial \xi}\right)_{x_B}$ is proportional to the driving force of the internal reaction, i.e., $\left(\frac{\partial G_m^{liq}}{\partial \xi}\right)_{x_B} \propto \left[\upsilon_A \mu_A^{liq}(x_B, \xi) + \upsilon_B \mu_B^{liq}(x_B, \xi) - \mu_{A_{\upsilon_A}B_{\upsilon_B}}^{liq}(x_B, \xi)\right]$ where $\mu_A^{liq}$ , $\mu_B^{liq}$ , and $\mu_{A_{\upsilon_A}B_{\upsilon_B}}^{liq}$ are chemical potentials of A, B, and $A_{\upsilon_A}B_{\upsilon_B}$ in the liquid phase. This equation is directly obtained from the variational derivative of the total free energy. At internal equilibrium, $\left(\frac{\partial G_m^{liq}}{\partial \xi}\right)_{x_B} = 0$. Note that the right-hand side of Eq. (3) can also include a gradient term of $\xi$ according to the conventional Allen-Cahn equations, as long as there is evidence that $\left(\frac{\partial G_m^{liq}}{\partial \xi}\right)_{x_B} = 0$ has multiple solutions, which correspond to different SRO states. If $\left(\frac{\partial G_m^{liq}}{\partial \xi}\right)_{x_B} = 0$ has only one solution, Eq. (3) will lead to a uniform distribution of $\xi$ under given $x_B$.

The evolution of the composition in liquid, $x_B$, is governed by

$$\frac{\partial x_B}{\partial t} = \nabla \cdot M \nabla \left(\frac{\partial G_m^{liq}}{\partial x_B}\right)_\xi \quad (4)$$

where $M$ is the interdiffusion mobility. Here we also neglect the gradient energy terms of $x_B$ unless there is evidence of phase separation. In reality, the evolution of $L_\xi$ can be much faster than that of $x_B$. Therefore, for each evolution step of $x_B$, Eq. (3) can either be evolved multiple times, or replaced by a Newton iteration scheme, until $\left(\frac{\partial G_m^{liq}}{\partial \xi}\right)_{x_B} = 0$ is achieved.

### *2.1.2 Phase-field model of liquid phases described by the MQMPA model*

Let us again consider a binary liquid with components A and B, and with the possibility of SRO. This time, we do not presume the stoichiometry of the SRO species; rather, we consider the quasichemical pair exchange reaction[2]

| $(A-A)_{pair} + (B-B)_{pair} = 2(A-B)_{pair}$ | (R2) |
|---|---|

with a Gibbs energy change of $\Delta g_{AB}$. The Gibbs free energy of the liquid solution is formulated as a function of the number of moles of A and B, $n_A$ and $n_B$, and the number of moles of A-A, B-B and A-B pairs, $n_{AA}$, $n_{BB}$ and $n_{AB}$, as[2]

| $G = n_A G_A^0 + n_B G_B^0 - T\Delta S^{conf} + \frac{n_{AB}}{2}\Delta g_{AB}$ | (5) |
|---|---|

Where $G_A^0$ and $G_B^0$ are molar Gibbs free energies of pure components A and B in liquid state, respectively, and $\Delta S^{conf}$ is the configurational entropy of mixing[2]:

| $\Delta S^{conf} = -R(n_A \ln X_A + n_B \ln X_B) - R\left(n_{AA}\ln\frac{X_{AA}}{Y_A^2} + n_{BB}\ln\frac{X_{BB}}{Y_B^2} + n_{AB}\ln\frac{X_{AB}}{2Y_AY_B}\right)$ | (6) |
|---|---|

Where $X_A$ and $X_B$ are mole fractions of A and B, $X_{AA}$, $X_{BB}$ and $X_{AB}$ are the mole fractions of the A-A, B-B and A-B pairs, respectively; $Y_A$ and $Y_B$ are coordination-equivalent fractions of A and B[2]:

| $X_i = \frac{n_i}{n_A + n_B} \quad (i = A\ or\ B)$ | (7a) |
|---|---|
| $X_{ij} = \frac{n_{ij}}{n_{AA} + n_{BB} + n_{AB}} \quad (i,j = A\ or\ B)$ | (7b) |
| $Y_i = \frac{Z_i n_i}{Z_A n_A + Z_B n_B} \quad (i = A\ or\ B)$ | (7c) |

Where $Z_i$ is the coordination number of A or B. The number of moles of the components and quasichemical pairs are not independent, constrained by the elemental balance equations[2]:

| | |
|---|---|
| $Z_A n_A = 2n_{AA} + n_{AB}$ | (8a) |
| $Z_B n_B = 2n_{BB} + n_{AB}$ | (8b) |

The coordinate numbers $Z_i$ can be composition-dependent[2]:

| | |
|---|---|
| $\frac{1}{Z_A} = \frac{1}{Z_{AA}^A}\left(\frac{2n_{AA}}{2n_{AA} + n_{AB}}\right) + \frac{1}{Z_{AB}^A}\left(\frac{n_{AB}}{2n_{AA} + n_{AB}}\right)$ | (9a) |
| $\frac{1}{Z_B} = \frac{1}{Z_{BB}^B}\left(\frac{2n_{BB}}{2n_{BB} + n_{AB}}\right) + \frac{1}{Z_{AB}^B}\left(\frac{n_{AB}}{2n_{BB} + n_{AB}}\right)$ | (9b) |

Where $Z_{AA}^A$ and $Z_{AB}^A$ are the coordination number of A when all its nearest neighbors are As and Bs, respectively; similarly, $Z_{BB}^B$ and $Z_{AB}^B$ are the coordination number of B when all its nearest neighbors are Bs and As, respectively. Furthermore, $\Delta g_{AB}$ can be formulated as a polynomial of pair fractions[2]:

| | |
|---|---|
| $\Delta g_{AB} = \Delta g_{AB}^0 + \sum_{i \geq 1} g_{AB}^{i0} (X_{AA})^i + \sum_{j \geq 1} g_{AB}^{0j} (X_{BB})^i$ | (10) |

With the equations above, for one mole of total components, i.e., $n_A + n_B = 1\ mole$, the molar Gibbs free energy of the liquid phase is a function of two independent variables: $n_B$ and $n_{AB}$ (see Appendix). Upon internal equilibrium, we should have $\left(\frac{\partial G}{\partial n_{AB}}\right)_{n_B} = 0$, indicating $n_{AB}$ can be treated as an IPOP to represent the extent of the internal quasichemical pair exchange reaction, i.e., $\xi = n_{AB}$. The evolution equation of $\xi$ can then be formulated similar to Eq. (3).

### 2.2 Phase-field model of a liquid phase and a stoichiometric solid phase

Let us now consider a liquid phase and a stoichiometric solid phase $\mathrm{A}_{\upsilon_A'}\mathrm{B}_{\upsilon_B'}$ in a binary A-B system where $\upsilon_A'$ and $\upsilon_B'$ are stoichiometric coefficients. To distinguish the two phase, as in [5], we introduce a phase order parameter $\eta$ to describe the extent of the stoichiometric reaction

$$\upsilon_A'\mathrm{A} + \upsilon_B'\,\mathrm{B} = \mathrm{A}_{\upsilon_A'}\mathrm{B}_{\upsilon_B'} \quad \text{(R3)}$$

$\eta = 0$ represents the liquid phase while $\eta = 1$ represents the solid phase.

The total free energy of the system, *F*, is formulated as

$$F = \int_V (g_{bulk} + g_{int})dV \quad (13)$$

where the bulk chemical free energy density $g_{bulk}$ is formulated as

$$g_{bulk} = \frac{1}{V_m}\{[1 - h(\eta)]G_m^{liq}(x_B, \xi) + h(\eta)G_m^{cmpd}\} \quad (14)$$

where $V_m$ is the molar volume of the system (assuming a constant), $h(\eta) = 6\eta^5 - 15\eta^4 + 10\eta^3$ is an interpolation function, $G_m^{cmpd}$ is the molar Gibbs free energy of the compound phase, which is a constant under constant temperature and pressure.

The interfacial free energy density is formulated as

$$g_{int} = w\eta^2(1 - \eta)^2 + \frac{1}{2}\kappa(\nabla\eta)^2 \quad (15)$$

where *w* is the double-well hump height, $\kappa$ is the gradient coefficient, both of which can be calculated from the interfacial energy and width. If the anisotropic interfacial energies need to be considered, then $\kappa$ and *w* can be made orientation-dependent. Note there is no gradient term of the composition[5].

The evolution equations of these phase-field variables will be formulated as

| | |
|---|---|
| $$\frac{\partial \eta}{\partial t} = -L_\eta \frac{\delta F}{\delta \eta} = -\frac{L_\eta}{V_m}\frac{\partial h}{\partial \eta}\left[G_m^{cmpd} - \upsilon_A' \mu_A^{liq}(x_B, \xi) - \upsilon_B' \mu_B^{liq}(x_B, \xi)\right] - L_\eta[w(4\eta^3 - 6\eta^2 + 2\eta) - \kappa \nabla^2 \eta]$$ | (16a) |
| $$\frac{\partial x_B}{\partial t} = \nabla \cdot M \nabla \left(\frac{\partial G_m^{liq}}{\partial x_B}\right)_\xi - \frac{\partial \left[h(\eta)\left(\frac{\upsilon_B'}{\upsilon_A' + \upsilon_B'} - x_B\right)\right]}{\partial t}$$ | (16b) |
| $$\frac{\partial \xi}{\partial t} = -L_\xi \frac{\delta F}{\delta \xi} = -L_\xi [1 - h(\eta)]\left(\frac{\partial G_m^{liq}}{\partial \xi}\right)_{x_B}$$ | (16c) |

where $L_\eta$ is a kinetic coefficient related to the reaction rate of (R3).

### 2.3 Numerical treatments

The evolution equations of the phase-field variables can be solved using different numerical algorithms including the finite difference (FD) method, the finite element method and the Fourier spectral method. We will primarily use the FD approach in this study. Besides, an important and realistic numerical issue one may encounter when directly incorporating the CALPHAD free energy functions is that these free energies involve logarithm functions, which causes convergence issues when the compositions and/or site fractions are higher than 1 or lower than 0. This issue becomes especially prominent for problems involve composition change, such as solidification and precipitation. There have been different numerical treatments to avoid such issues. To improve the numerical convergence while maintaining the mass conservation, we replace the Allen-Cahn equation for IPOP evolution with a Newton iteration scheme,

| | |
|---|---|
| $$\frac{\partial \xi}{\partial t} = -L_\xi \left(\frac{\partial G_m^{liq}}{\partial \xi}\right)_{x_B} \Bigg/ \left(\frac{\partial^2 G_m^{liq}}{\partial \xi^2}\right)_{x_B}$$ | (17) |

where $L_\xi$ now become an adjustable parameter to ensure the convergence of the iteration. The iteration can be designed to stop either when the $\left(\frac{\partial G_m^{liq}}{\partial \xi}\right)_{x_B}$ value is lower than certain threshold ($10^{-7}$) for simulations under the internal equilibrium of $\xi$, or after a certain number of iteration steps, for general nonequilibrium conditions of $\xi$.

## 3. Worked Examples

To test the validity of these phase-field models, we apply the models to the liquid phases described by different thermodynamic models in realistic materials systems. For each example below, we will first perform single-phase, uniform simulations by only evolving the IPOP in the liquid phase under fixed composition, to compare with thermodynamic equilibrium calculations in terms of compositions, IPOPs and site fractions. We will then perform one-dimensional (1-D) simulations of diffusion-controlled growth of a stoichiometric solid in the liquid phase.

### *3.1 Na-Sb system: associate model for liquid*

The Na-Sb alloy is a candidate for anodes of sodium-ion batteries[6]. The microstructure of the alloy is critical to the battery performance. The ($Na$+$Na_3Sb$) two-phase microstructure can be achieved and controlled via solidification. Therefore, the thermodynamic and kinetic behaviors of the liquid phase in Na-Sb are of great research interest. In [7], the liquid phase of the Na-Sb system is described by the sublattice model $(Na, Sb, Na_3Sb)_1$ with $Na_3Sb$ being the associate. Therefore, we select $\xi = y_{Na_3Sb}$ as the IPOP. We first examine the evolution of the IPOP at $T$=1000 K with a uniform $x_{Sb} = 0.1$ and an initial IPOP of $\xi_0 = 0.1$, as shown in Fig. 2(a). The phase-field simulation clearly evolves to the correct thermodynamic equilibrium $\xi = 0.142857$ (based on OpenCalphad calculations) with various $L_\xi$ values, which corresponds to the minimum $G_m^{liq}$ under $x_{Sb} = 0.1$. With this validation, we further evolve the IPOP under different uniform $x_{Sb}$ values ranging from 0 to 1, starting with $\xi_0 = 0.1\frac{x_{Sb}}{1-3x_{Sb}}$ for $x_{Sb} \leq 0.25$ and $\xi_0 = 0.1\frac{1-x_{Sb}}{3x_{Sb}}$ for $x_{Sb} > 0.25$. The evolution trajectories from $\xi_0$ to equilibrium are shown by the curves at different evolution steps in Fig. 2(b). Those equilibrium $\xi$ values are all consistent with OpenCalphad calculations. It should be noted that the physically meaningful $\xi$ values are confined to the regions of $0 < \xi < \frac{x_{Sb}}{1-3x_{Sb}}$ for $0 < x_{Sb} \leq 0.25$ and $0 < \xi < \frac{1-x_{Sb}}{3x_{Sb}}$ for $0.25 < x_{Sb} < 1$, so that all the sublattice site fractions lie between 0 and 1. The calculated equilibrium $\xi$ values are very close to the upper boundaries of these regions.

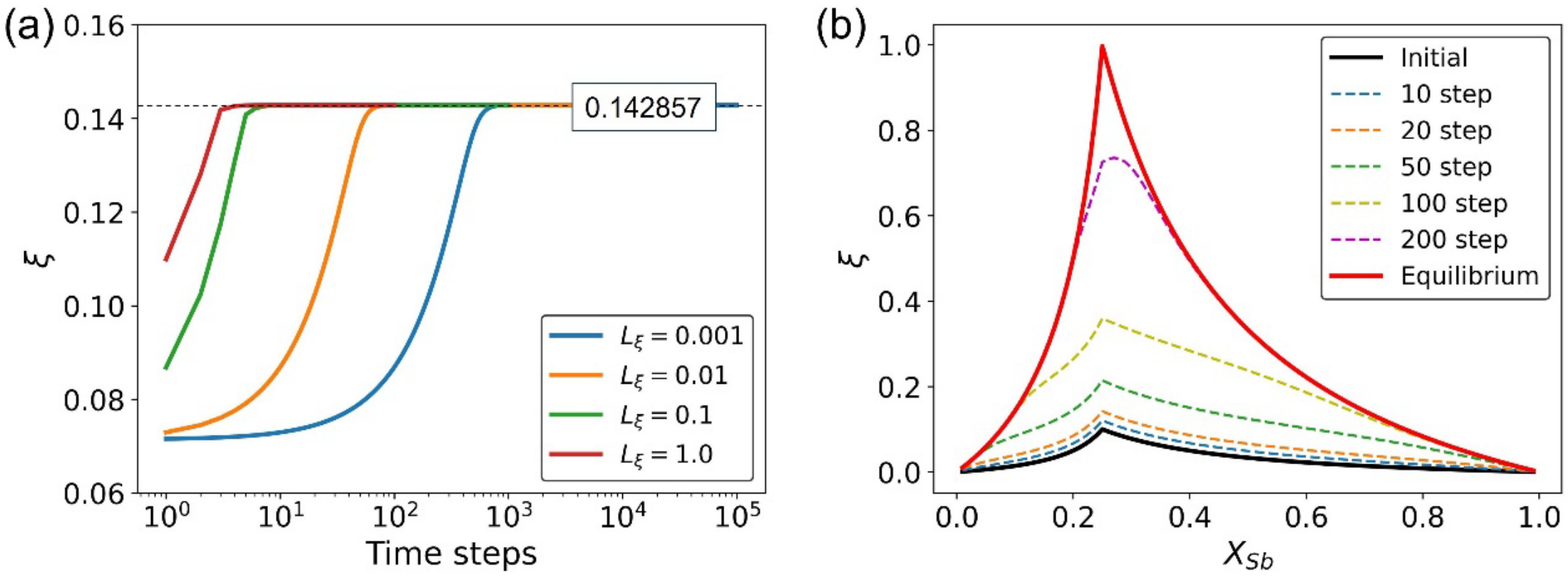

Fig. 2 Phase-field simulation of IPOP evolution in homogeneous liquid phases of Na-Sb alloy at 1000 K. (a) IPOP evolution under different $L_\xi$ values for a fixed Sb composition of 0.1 and initial IPOP value of 0.1, towards the equilibrium value of 0.142857; (b) IPOP evolution trajectories for different Sb compositions and initial IPOP values towards equilibrium (red line).

With this verification, we then apply the model to the (liquid+$Na_3Sb$) two phase system at $T$=1000 K, where $Na_3Sb$ is considered as a stoichiometric compound. The 1-D simulation region is uniformly discretized into 1024 grid points with a grid spacing of $\Delta x$ = 1 nm. At the beginning of the simulation, we put a small nucleus of $Na_3Sb$ with a length of 8 nm at the left-hand side of the simulation region. The rest of the simulation region is composed of supersaturated liquid phase with uniform $x_{Sb} = 0.1$ and $\xi = 0.1$. Zero-flux boundary conditions are used for all phase-field variables. The simulations are performed under fixed $L_\eta = 1$ and $M = 1$, with varying $L_\xi$ from 0 to 1. This set of simulations can capture different $Na_3Sb$ evolution pathways due to the competition between SRO in the liquid phase and the formation of second-phase $Na_3Sb$. When $L_\xi = 0$, the SRO is frozen-in (no evolution) and the $G_m^{liq}$ is described by the blue curve in Fig. 3(a); the common tangent construction gives the composition of $x_{Sb} = 0.076923$ in the liquid phase at two-phase equilibrium. In contrast, when $L_\xi = 1$, the SRO is much faster than diffusion and the second-phase formation, and the $G_m^{liq}$ is described by the yellow curve in Fig. 3(a), each point on which satisfies the internal equilibrium condition of $\left(\frac{\partial G_m^{liq}}{\partial \xi}\right)_{x_{Sb}} = 0$. The common tangent construction now gives a composition of $x_{Sb} = 0.050843$ and the SRO extent of $\xi = 0.059994$ in the liquid phase at the two-phase equilibrium.

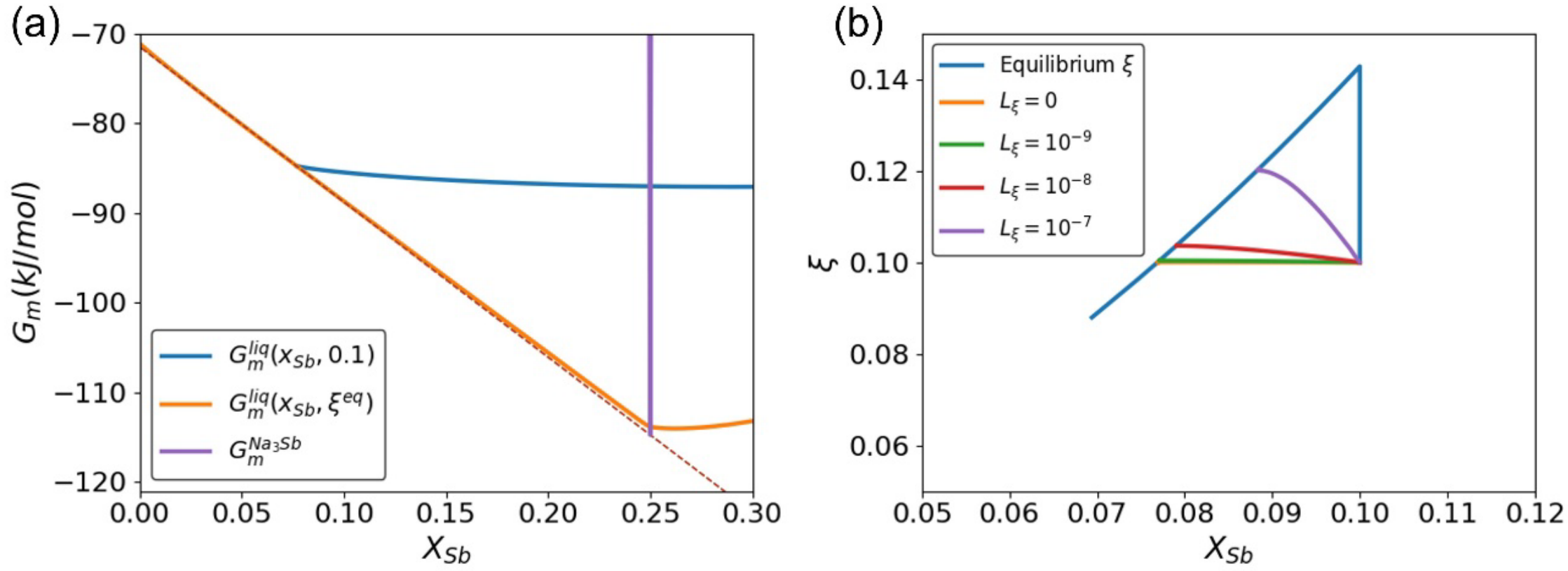

Fig. 3 (a) Free energy landscapes for the liquid and compound phases. (b) Evolution trajectories for $\xi$ and $x_{Sb}$ under different $L_\xi$.

For $L_\xi$ values between these two extremes, it is expected that the competition between SRO in the liquid phase and the formation of second-phase $Na_3Sb$ becomes more complicated. To quantify the kinetic pathways, we track the average $x_{Sb}$ and $\xi$ values in the liquid phases during evolution, as plotted in Fig. 3(b). As $L_{\xi}$ increases, the evolution trajectory changes from the horizontal line (yellow) to the $L_{\xi} = 1$ case (blue line). The detailed evolution profiles of all the phase-field variables for the $L_{\xi} = 1$ case (assuming internal equilibrium $\xi$ at every time step) are shown in Fig. 4.

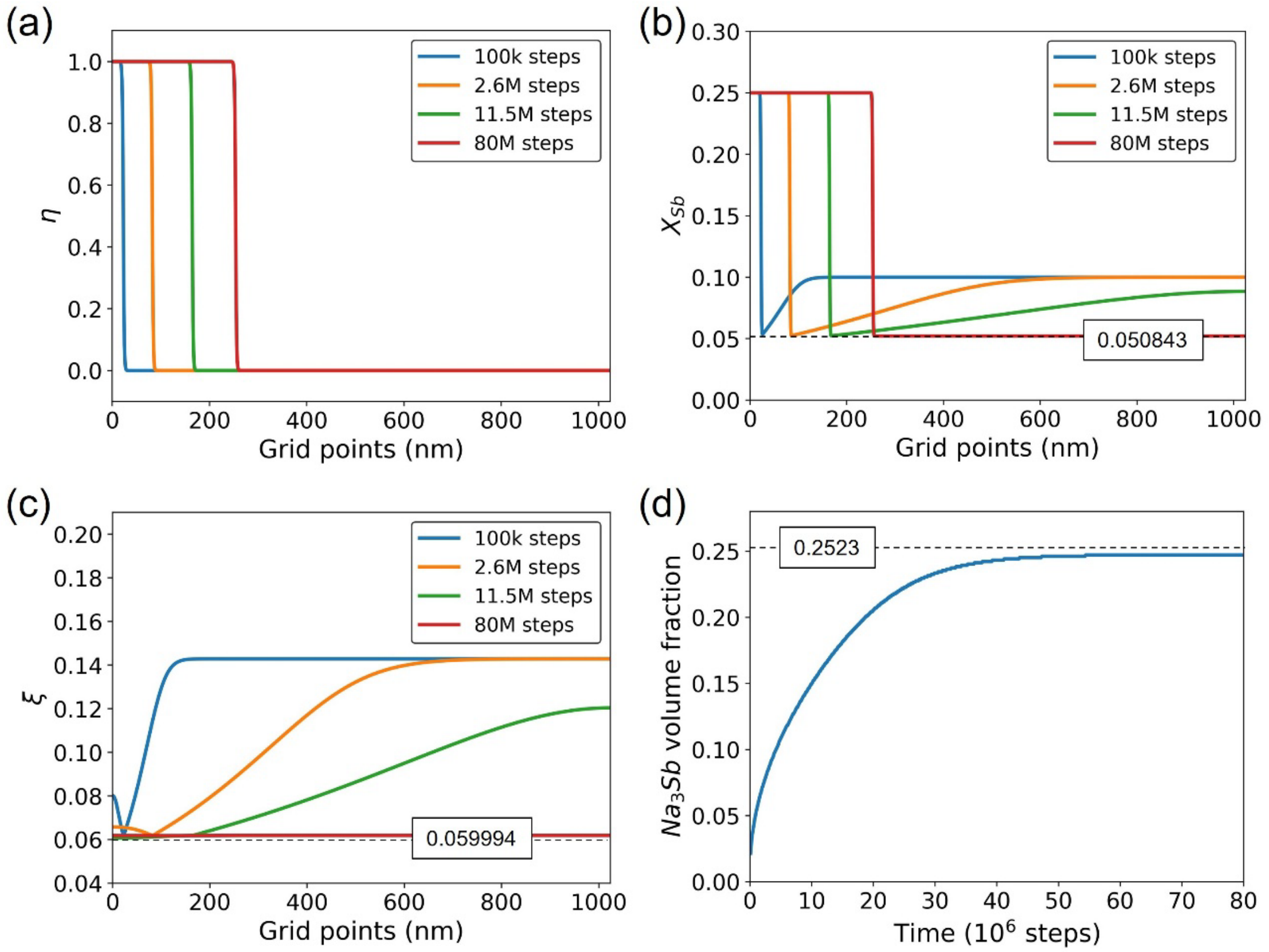

Fig. 4 Phase-field simulation of stoichiometric $Na_3Sb$ growth from a liquid phase in a Na-10 at.%Sb alloy with initial $\xi = 0.1$ at 1000 K and $\xi$ reaching internal equilibrium for every time step. (a) Evolution of the phase order parameter. (b) Evolution of the Sb composition. (c) Evolution of the IPOP. (d) Evolution of the volume fraction of the $Na_3Sb$ phase.

### *3.2 Na-Sn system: MQMPA for liquid*

Similar to Na-Sb, the Na-Sn alloy is also a candidate for anodes of sodium-ion batteries[6]. Compared with Na-Sb, the Na-Sn system involves more stoichiometric compounds, indicating more complicated ordering behaviors between Na and Sn. As in[8], the liquid phase of the Na-Sn system is described by MQMPA. Therefore, we select $\xi = n_{NaSn}$ as the IPOP. The simulation of IPOP evolution at $T$=600 K with a uniform $x_{Sn} = 0.1$ and an initial IPOP of 0.1 approaches the correct equilibrium value of $\xi$, as shown in Fig. 5.

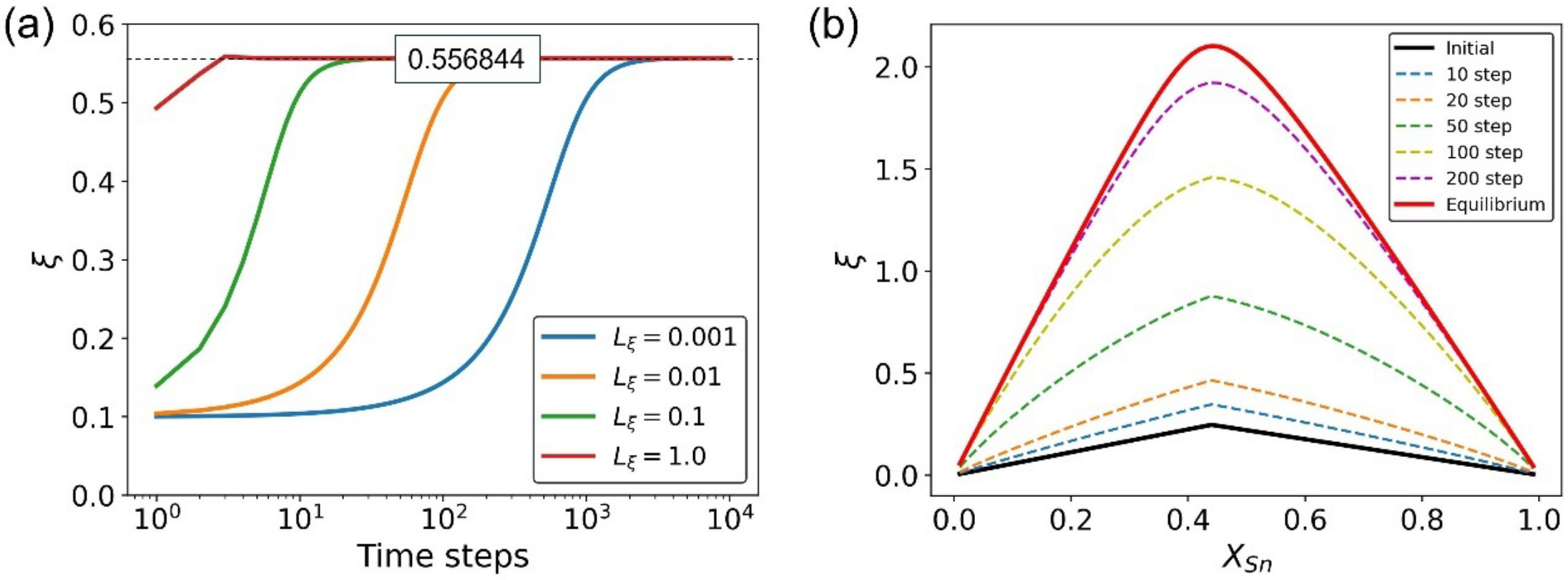

Fig. 5 Phase-field simulation of IPOP evolution in homogeneous liquid phases of Na-Sn alloy at 600 K. (a) IPOP evolution under different $L_\xi$ values for a fixed Sn composition of 0.1 and initial IPOP value of 0.1, towards the equilibrium value of 0.556844; (b) IPOP evolution trajectories for different Sn compositions and initial IPOP values towards equilibrium (red line).

Since there are multiple solid stoichiometric compounds in the Na-Sn system, we perform simulations only for diffusion-controlled growth of $Na_{15}Sn_4$, as shown in Fig. 6.

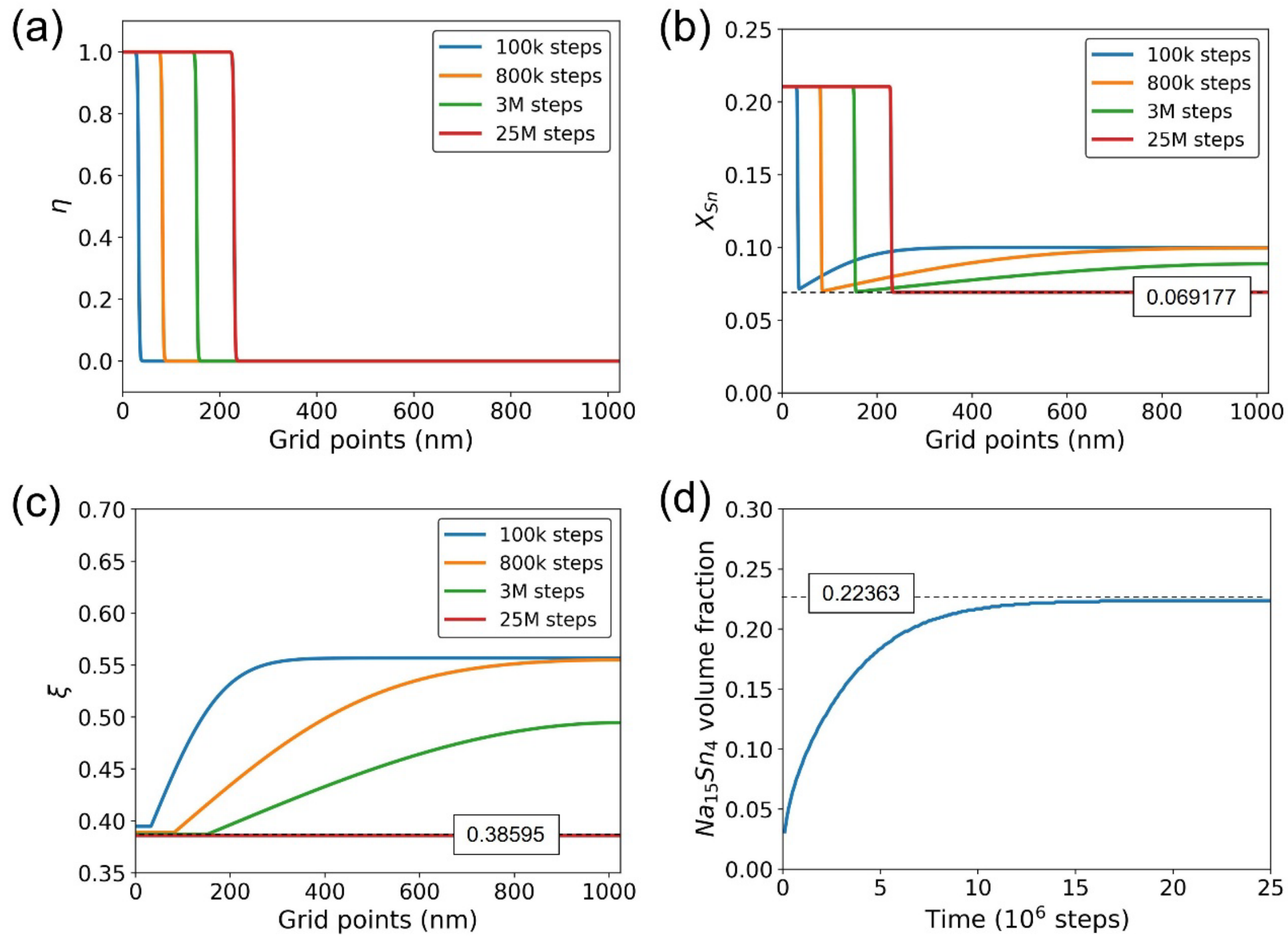

Fig. 6 Phase-field simulation of stoichiometric $Na_{15}Sn_4$ growth from a liquid phase in a Na-10 at.%Sn alloy with initial $\xi = 0.1$ at 600 K and $\xi$ reaching internal equilibrium for every time step. (a) Evolution of the phase order parameter. (b) Evolution of the Sn composition. (c) Evolution of the IPOP. (d) Evolution of the volume fraction of the $Na_{15}Sn_4$ phase.

## 4. Conclusion

In this work, we build a direct seamless connection between the CALPHAD thermodynamic modeling of the liquid phases with ordering, using either the associate model or the MQMPA, and the phase-field simulations involving such liquid phases. This connection is achieved by identifying the internal processes within each thermodynamic description of the liquid phase, defining the IPOPs to describe the extent of these internal processes, and evolving these IPOPs using Allen-Cahn equations, which can be coupled with other phase-field evolution equations. The developed phase-field schemes are validated against thermodynamic equilibrium calculations in Na-Sb and Na-Sn systems. This connection enables the accurate phase-field simulation of microstructure evolution during solidification and the exploration of how the SRO processes can affect the microstructure evolution kinetics, providing better understanding and control of microstructure evolution in liquids with negative enthalpy of mixing.

**Acknowledgement**

This work is supported by the start-up fund and the College of Engineering Strategic Research Initiative Grant from The Ohio State University. The computation is supported by Ohio Supercomputer Center (OSC).

**Appendix: Molar Gibbs free energy of liquid phase in MQMPA**

There are many variables involved in the MQMPA formulation. Based on their relationships in Eqs. (), it can be shown that there are only 3 independent variables, $n_A$, $n_B$ and $n_{AB}$, and all the other variables can be expressed in terms of these 3 variables as listed below:

$$n_{AA} = \frac{1}{2} Z_{AA}^{A} \left( n_A - \frac{n_{AB}}{Z_{AB}^{A}} \right)$$

$$n_{BB} = \frac{1}{2} Z_{BB}^{B} \left( n_B - \frac{n_{AB}}{Z_{AB}^{B}} \right)$$

$$Z_A = Z_{AA}^{A} + \frac{n_{AB}}{n_A} \left( 1 - \frac{Z_{AA}^{A}}{Z_{AB}^{A}} \right)$$

$$Z_B = Z_{BB}^{B} + \frac{n_{AB}}{n_B} \left( 1 - \frac{Z_{BB}^{B}}{Z_{AB}^{B}} \right)$$

$$X_{AA} = \frac{Z_{AA}^{A} Z_{AB}^{B} (n_{AB} - n_{A} Z_{AB}^{A})}{n_{AB}(Z_{AA}^{A} Z_{AB}^{B} + Z_{AB}^{A} Z_{BB}^{B} - 2 Z_{AB}^{A} Z_{AB}^{B}) - Z_{AB}^{A} Z_{AB}^{B} (n_{A} Z_{AA}^{A} + n_{B} Z_{BB}^{B})}$$

$$X_{AB} = \frac{-2 n_{AB} Z_{AB}^{A} Z_{AB}^{B}}{n_{AB}(Z_{AA}^{A} Z_{AB}^{B} + Z_{AB}^{A} Z_{BB}^{B} - 2 Z_{AB}^{A} Z_{AB}^{B}) - Z_{AB}^{A} Z_{AB}^{B} (n_{A} Z_{AA}^{A} + n_{B} Z_{BB}^{B})}$$

$$X_{BB} = \frac{Z_{AB}^{A} Z_{BB}^{B} (n_{AB} - n_{B} Z_{AB}^{B})}{n_{AB}(Z_{AA}^{A} Z_{AB}^{B} + Z_{AB}^{A} Z_{BB}^{B} - 2 Z_{AB}^{A} Z_{AB}^{B}) - Z_{AB}^{A} Z_{AB}^{B} (n_{A} Z_{AA}^{A} + n_{B} Z_{BB}^{B})}$$

$$Y_{A} = \frac{-Z_{AB}^{B} \left( n_{A} Z_{AA}^{A} Z_{AB}^{A} + n_{AB} (Z_{AB}^{A} - Z_{AA}^{A}) \right)}{n_{AB}(Z_{AA}^{A} Z_{AB}^{B} + Z_{AB}^{A} Z_{BB}^{B} - 2 Z_{AB}^{A} Z_{AB}^{B}) - Z_{AB}^{A} Z_{AB}^{B} (n_{A} Z_{AA}^{A} + n_{B} Z_{BB}^{B})}$$

$$Y_{B} = \frac{-Z_{AB}^{A} \left( n_{B} Z_{AB}^{B} Z_{BB}^{B} + n_{AB} (Z_{AB}^{B} - Z_{BB}^{B}) \right)}{n_{AB}(Z_{AA}^{A} Z_{AB}^{B} + Z_{AB}^{A} Z_{BB}^{B} - 2 Z_{AB}^{A} Z_{AB}^{B}) - Z_{AB}^{A} Z_{AB}^{B} (n_{A} Z_{AA}^{A} + n_{B} Z_{BB}^{B})}$$

**Conflict of interest statement**

On behalf of all authors, the corresponding author states that there is no conflict of interest.

**Data availability statement**

Data sets generated during the current study are available from the corresponding author on reasonable request.